\documentclass[fleqn,usenatbib,useAMS]{mnras}

\usepackage{graphicx,times}
\usepackage{subfigure}
\newcommand{\be}{\begin{equation}}
\usepackage{threeparttable}
\usepackage{booktabs}
\newcommand{\ee}{\end{equation}}
\newcommand{\bea}{\begin{eqnarray}}
\newcommand{\eea}{\end{eqnarray}}

\usepackage{enumerate}
\usepackage{amsmath}
\usepackage{cases}
\usepackage{longtable}
\usepackage{hyperref}
\usepackage{epstopdf}
\usepackage{amsmath,bm}
\usepackage{amssymb}
\usepackage{natbib}
\usepackage{morefloats}
\usepackage{multirow}
\usepackage{array}
\usepackage{verbatim}

\title[Projected velocity statistics]{Projected velocity statistics of interstellar turbulence }

\author[Siyao Xu]{
Siyao Xu\thanks{E-mail: sxu93@wisc.edu }
\\
Department of Astronomy, University of Wisconsin, 475 North Charter Street, Madison, WI 53706, USA; Hubble Fellow\\
}

\begin{document}
\label{firstpage}
\pagerange{\pageref{firstpage}--\pageref{lastpage}}
\maketitle

\begin{abstract}
Velocity statistics is a direct probe of the dynamics of interstellar turbulence. 
Its observational measurements are very challenging due to the convolution between density and velocity 
and projection effects.  
We introduce the projected velocity structure function, which can be generally applied to statistical studies of both
sub- and super-sonic turbulence in different interstellar phases. 
{It recovers the turbulent velocity spectrum from the projected velocity field 
in different regimes, 
and when the thickness of a cloud is less than the driving scale of turbulence, 
it can also be used to determine the cloud thickness and the turbulence driving scale.} 
By applying it to the existing core velocity dispersion measurements of the Taurus cloud, 
we find a transition from the Kolmogorov to the Burgers scaling of turbulent velocities with decreasing length scales, 
corresponding to the large-scale solenoidal motions and small-scale compressive motions, respectively. 
The latter occupy a small fraction of the volume and 
can be selectively sampled by clusters of cores with the typical cluster size indicated by the transition scale. 
\end{abstract}

\begin{keywords}
turbulence -- methods: statistical -- ISM: clouds
\end{keywords}

\section{Introduction}

Turbulence exists over an enormous range of length scales 
\citep{Armstrong95,Sca04,CheL10}. 
Its solenoidal and compressional components play different roles in 
affecting the gas dynamics, density and magnetic field distribution 
(e.g., \citealt{LP00,Burk09,Pan10,XuZ16,XuZ17,Yu17,Moc18,XuJ18}),
and associated important physical processes in the interstellar medium (ISM), such as the 
dynamo amplification of magnetic fields
\citep{XL16,XuG19},
reconnection diffusion of magnetic fields 
\citep{Sant10,Xurd19},
cosmic ray propagation 
\citep{Xuc16,XLcr18},
and star formation 
\citep{Mckee_Ostriker2007,Fed12}.

To study the dynamics of interstellar turbulence and identify its solenoidal and compressional components on different length scales, 
statistical measurements of turbulent velocities are urgently needed,
which, however, are very challenging in observations due to the complex convolution between density and velocity. 
Because of this difficulty, even the simplest velocity statistics, i.e., the velocity spectrum in Fourier space and velocity structure function in 
real space, have been scarcely measured 
\citep{Laz09rev,Chep10}.
\footnote{We note that the spectral and structure function measurements of turbulent velocities are different from the 
linewidth-size measurements
\citep{Lars81}. 
The former are not limited to particular turbulent motions in particular interstellar phases, 
whereas the latter reflect the internal turbulent motions of dense structures in cold interstellar phases.}
To disentangle the statistics of velocity and density, different techniques have been developed, 
including 
the Velocity Channel Analysis (VCA)
\citep{LP00},
the Velocity Coordinate Spectrum (VCS)
\citep{LP06,Laz09rev},
principal component analysis (PCA)
\citep{Hey97},
modified velocity centroids
\citep{LE03},
and centroid velocity increments 
\citep{Lis96}.
More recently, the core velocity dispersion (CVD) technique was proposed by 
\citet{Qi12}. 
By using the peak velocity of the emission profiles of discrete dense cores, 
the CVD measurement by itself is less prone to density fluctuations
and can more cleanly extract the velocity statistics in molecular clouds (MCs).

Statistical measurements of turbulent velocities are also subject to projection effects. 
To fully and properly exploit the available information in observations and compare it with theoretical models of turbulence, 
the knowledge of the scaling relation between the 2D and 3D statistics is essential. 
The structure function of projected variables developed by 
\citet{LP16} (hereafter LP16)
shows that the 2D-to-3D conversion depends on 
(a) the relation of 
the correlation length of variable fluctuations
to the thickness of the observed volume along the line of sight (LOS),
(b) the steepness of the Fourier spectrum of variable fluctuations, and 
(c) the length scale of interest. 
It has been tested with synthetic observations
\citep{Lee16,Jia18}
and applied to interpreting the interstellar Faraday rotation measurements
\citep{XuZ16}.

To observationally obtain the statistics and dynamics of interstellar turbulence, 
we aim to construct the structure function of projected turbulent velocities, which is described in 
\S 2. 
By applying it to analyzing 
the CVD measurements of the Taurus cloud, we study the statistical properties and dynamics of 
supersonic turbulence in MCs (\S 3 and \S 4). 
The conclusions are in \S 5.

\section{Structure function of projected turbulent velocities}
\label{sec:sf}

The correlation function (CF) and 
the structure function (SF) 
are frequently used to describe the velocity statistics in a turbulent medium 
\citep{LP00,LP04,LP06}. 
Under the assumption that the turbulent velocity field is homogeneous and isotropic,
\footnote{Although the turbulent velocity in the presence of magnetic field has scale-dependent anisotropy 
in the local frame aligned with the locally averaged magnetic field direction 
\citep{GS95,LV99,CLV_incomp},
isotropic statistics are applicable to observational data in the global frame aligned with the large-scale mean magnetic field 
\citep{Esq03}.
For the purpose to study turbulence anisotropies, the velocity correlation tensors of different MHD modes were discussed in 
\citet{Kan16}.}
the CF of the LOS component of turbulent velocity $u_z$ is 
\begin{equation}
    \xi(R,\Delta z) = \langle u_z(\bm{X_1},z_1) u_z(\bm{X_2},z_2)\rangle,
\end{equation}
where $R = |\bm{X_1} - \bm{X_2}|$, $\Delta z = z_1 - z_2$, 
$\bm{X}$ and $z$ represent the position on the plane of sky and the distance along the LOS, 
and $\langle...\rangle$ denotes the ensemble average. 
The SF of $u_z$ is defined as 
\begin{equation}
      d(R, \Delta z) = \langle [u_z(\bm{X_1},z_1) - u_z(\bm{X_2},z_2) ]^2 \rangle.
\end{equation}
When both the outer and inner cutoffs of power-law spectrum of velocity fluctuations are explicitly given,
the CF and SF are related by 
\citep{LP06},
\begin{equation}\label{eq: relcfsf}
\begin{aligned}
   &   d(R, \Delta z)  = 2 [\xi(0) - \xi(R,\Delta z)],   \\
   &   \xi(R,\Delta z) = \frac{1}{2} [d(\infty) - d(R, \Delta z)].
\end{aligned}
\end{equation}

To describe the scaling properties of CF and SF, 
we adopt the power-law correlation model presented in 
LP16
and have 
\begin{equation}\label{eq: cfvz}
      \xi(R,\Delta z) =  \langle u_z^2 \rangle   \frac{L_i^m}{L_i^m  +  (R^2 + \Delta z^2)^\frac{m}{2}} ,
\end{equation}
with 
\begin{equation}
    \xi(0) = \langle u_z^2 \rangle ,
\end{equation}
and 
\begin{equation}
      d(R, \Delta z)  = 2\langle u_z^2 \rangle \frac{(R^2 + \Delta z^2)^\frac{m}{2}}{L_i^m  +  (R^2 + \Delta z^2)^\frac{m}{2}}  ,
\end{equation}
which satisfy the relation in Eq. \eqref{eq: relcfsf}.
Here $L_i$ is the correlation length of turbulent velocities, that is, the energy injection scale of turbulence. 
The above expressions of  $\xi(R,\Delta z)$ and $d(R, \Delta z)$ can be used on length scales both larger and smaller than $L_i$. 
For both sub- and super-sonic turbulence, the turbulent velocities are dominated by large-scale fluctuations and has 
a steep spectrum
\citep{CL03}.
Therefore, 
the 3D spectral index $\alpha$ of turbulent velocities is related to the power-law index $m$ by 
\begin{equation}\label{eq: ressf}
    \alpha = -m - 3.
\end{equation}

In observations, 
we have the turbulent velocities integrated over the LOS in a cloud with the LOS thickness $L$. 
The projected velocity SF is 
\begin{equation}\label{eq: drpjt}
\begin{aligned}
   D(R) &= \langle [u_z(\bm{X_1}) - u_z(\bm{X_2}) ]^2 \rangle   \\
                    &= \Big\langle \Big[ \int_0^L dz u_z(\bm{X_1},z)   - \int_0^L dz u_z(\bm{X_2},z)  \Big] ^2 \Big\rangle    \\
               & = 4  \langle u_z^2 \rangle \int_0^L d\Delta z (L-\Delta z)  \\
               & ~~~~~~ \Bigg[    \frac{L_i^m}{L_i^m  +  \Delta z^m} -    \frac{L_i^m}{L_i^m  +  (R^2 + \Delta z^2)^\frac{m}{2}}  \Bigg] ,
\end{aligned}
\end{equation}
where the expression in Eq. \eqref{eq: cfvz} is used. 
$D(R)$ has different asymptotic forms in different ranges of length scales. 
In the case of a ``thin" cloud with $L < L_i$, there is 
\begin{subnumcases}
     { D(R) \approx  \label{eq: drtic}}
       4  \langle u_z^2 \rangle L_i^{-m}LR^{m+1}, ~~~~~~R<L,  \label{eq: thinsa1}\\   
       2  \langle u_z^2 \rangle  L_i^{-m}L^2 R^m,~~~~ L<R<L_i,  \label{eq: thinsa2}\\      
       2  \langle u_z^2 \rangle L^2, ~~~~~~~~~~~~~~~~~~~~~~ R > L_i.  \label{eq: thinsa3}
\end{subnumcases}
In the case of a ``thick" cloud with $L>L_i$, we find 
\begin{equation}\label{eq: drthc}
    D(R) \approx 4  \langle u_z^2 \rangle L_i^{-m} L R^{m+1}, ~~~~~~~~R<L_i.
\end{equation}
{The detailed derivation is provided in Appendix \ref{sec:app}. 
In Fig. \ref{fig: illu}, we compare the above approximate expressions with the numerically calculated $D(R)$. 
We adopt the Kolmogorov scaling of turbulent velocities ($m = 2/3$) as an illustration.}
The scaling relation of $D(R)$ to $d(R,\Delta z)$ depends on the 
relation between $L$ and $L_i$ and the length scale of interest. 
We note that only in the case of $L < R < L_i$, 
do the scaling exponents of $D(R)$ and $d(R,\Delta z)$ coincide. 
{When $R < L$ for both thin and thick clouds (Eqs. \eqref{eq: thinsa1} and \eqref{eq: drthc}), 
the projection effect becomes important, and thus 
the 2D SF has a steeper slope than that of the 3D SF. 
In the case of a thin cloud, by measuring the 2D velocity SF over a broad range of length scales, 
we are able to determine $L$ and $L_i$ as the transition scales between different regimes (see Fig. \ref{fig: thin}).}

\begin{figure*}
\centering
\subfigure[Thin cloud ($L < L_i$)]{
   \includegraphics[width=8cm]{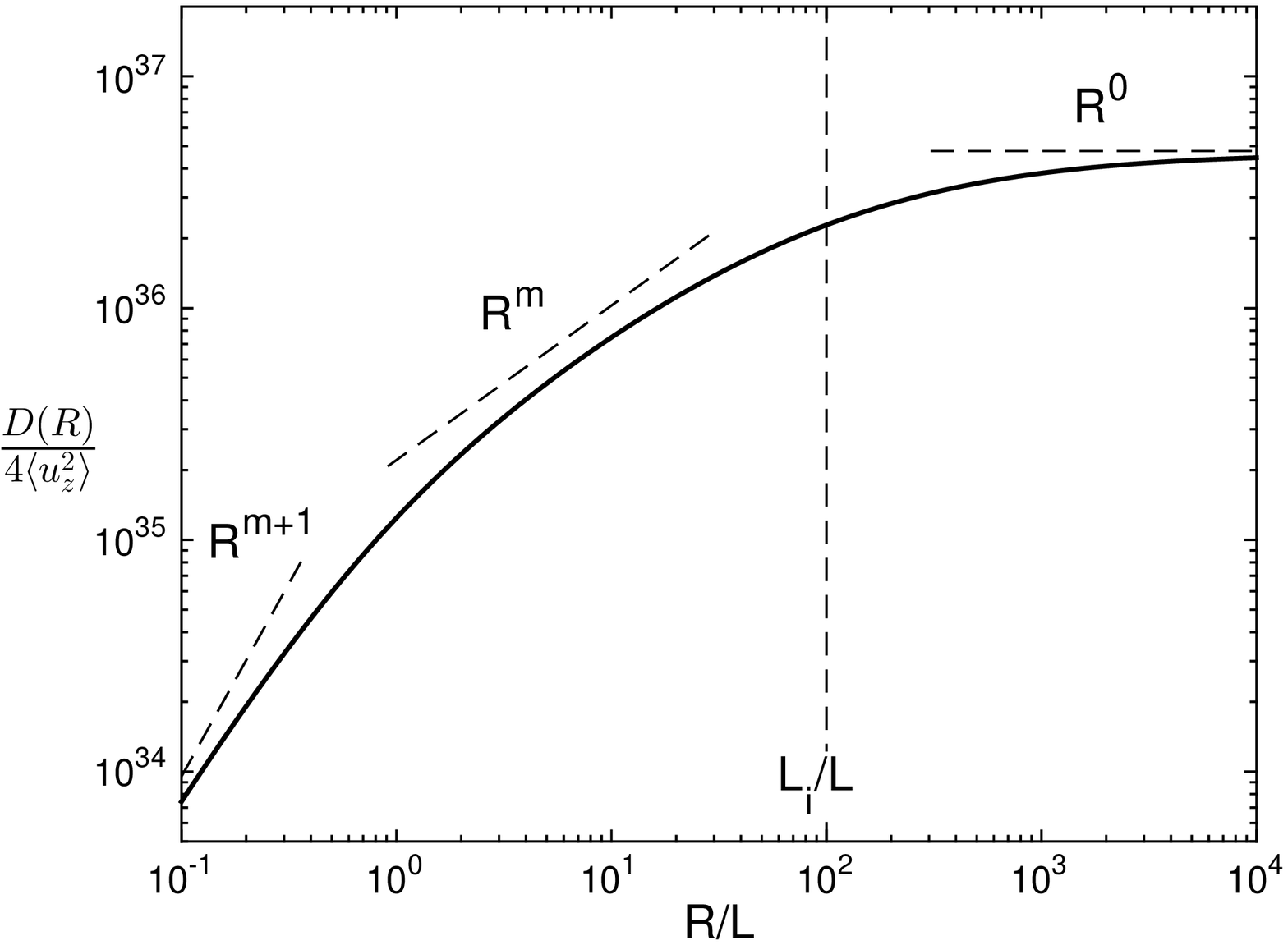}\label{fig: thin}}
\subfigure[Thick cloud ($L > L_i$)]{
   \includegraphics[width=8cm]{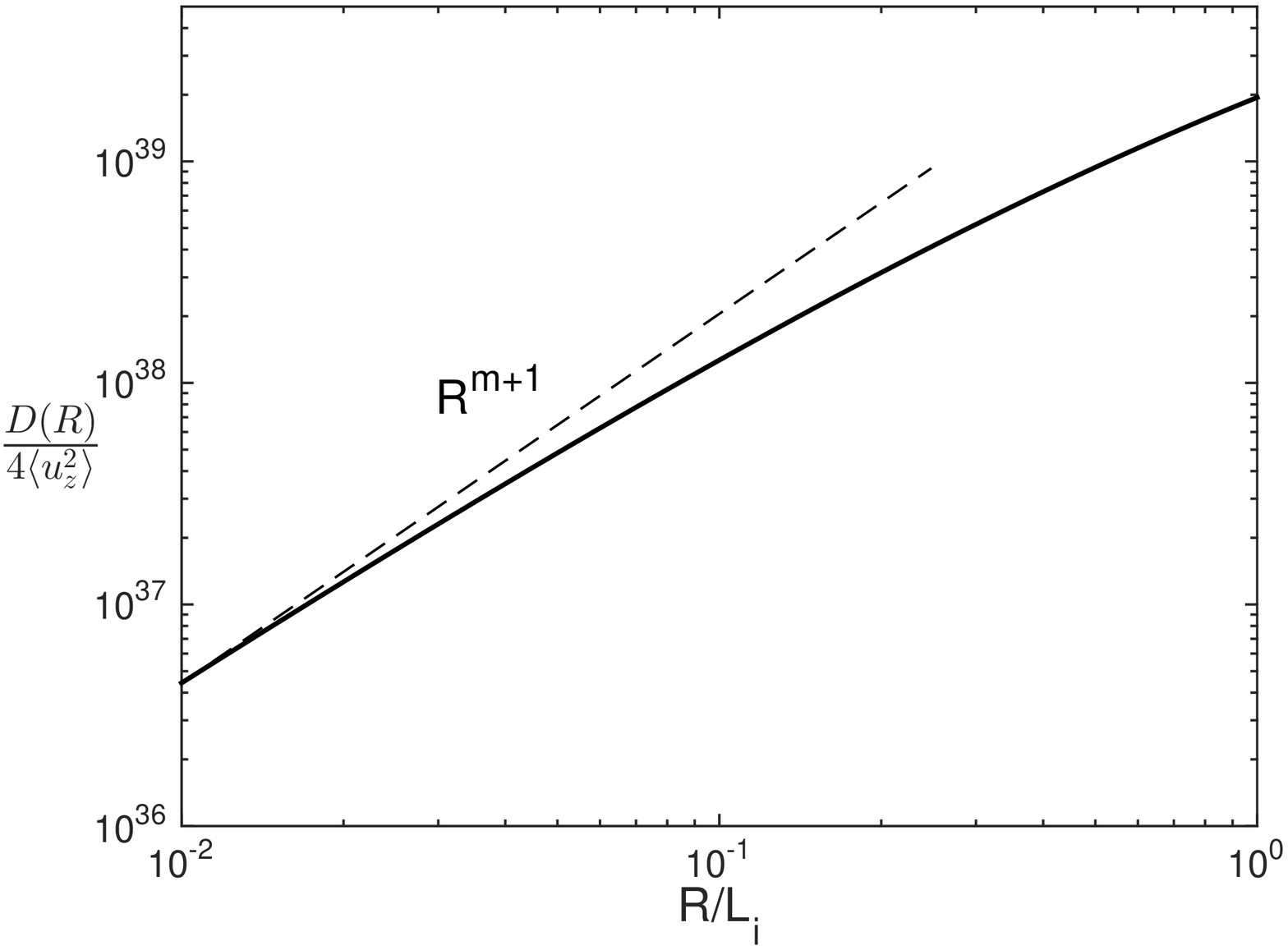}\label{fig: thick}}
\caption{{Analytical approximations of $D(R) / 4 \langle u_z^2 \rangle$ (dashed lines) in comparison with the numerically calculated results (solid lines) in different regimes.   
The dashed lines correspond to Eqs. \eqref{eq: thinsa1}, \eqref{eq: thinsa2}, \eqref{eq: thinsa3} in (a), 
and Eq. \eqref{eq: drthc} in (b). 
The vertical dashed line in (a) indicates the value of $L_i$ adopted in the case of a thin cloud. 
}}
\label{fig: illu}
\end{figure*}

The SF of projected physical variables was earlier introduced by 
LP16 for rotation measure fluctuations and later applied by 
\citet{XuZ16}
to explaining the observed Faraday rotation measurements in the ISM. 
Besides rotation measure fluctuations and projected turbulent velocities considered here, 
it can be generally applied to statistical studies of 
projected observables related to interstellar turbulence, 
for both sub- and super-sonic turbulence in different interstellar phases.

\section{Scalings of supersonic turbulence }

\subsection{Scalings of turbulent velocities}
\label{ssec: sctv}

The turbulent velocities in supersonic turbulence can be decomposed into
solenoidal and compressional components. 
At the energy equipartition between the two components, 
the fraction of turbulent energy in compressional component saturates at $1/3$ at a large sonic Mach number 
\citep{Pan10}.
The power-law indices of the turbulent velocity spectra of solenoidal and compressional components are close to 
$\alpha_s =- 11/3$ and $\alpha_c = - 4$, respectively
\citep{KowL10}. 
The former is known as the Kolmogorov spectrum of incompressible turbulence, 
and the latter is the Burgers spectrum of shock-dominated turbulence. 
We note that due to the limited inertial range of turbulence, 
the observed velocity scalings in numerical simulations can strongly depend on the numerical driving of turbulence. 
In addition, in most simulations with a barotropic equation of state,
the baroclinic effect is absent and thus 
the conversion from compressive to solenoidal motions is suppressed
\citep{Pad16}.
Hence, one must be cautious when numerically studying the scalings of turbulent velocities due to the above artifacts.

\subsection{Scalings of density fluctuations}
\label{ssec: sdf}

As shown by numerical experiments of solenoidally driven supersonic isothermal turbulence,
the probability density function of density fluctuations follows a lognormal distribution
\citep{PN02},
\begin{equation}
     p(\xi)  =  \frac{1}{\sqrt{2\pi\sigma^2}}\exp\Bigg[-\frac{(\xi-\mu)^2}{2\sigma^2}\Bigg],
\end{equation}
where 
\begin{equation}
  \xi = \ln \Big(\frac{\rho}{\langle\rho\rangle}\Big),  \sigma^2 = \ln (1+b^2 \langle M_s \rangle^2),
  \mu = -\frac{\sigma^2}{2},
\end{equation}
with $\langle\rho\rangle$ as the mean density, $\langle M_s \rangle$ as the rms sonic Mach number, 
and the constant $b$. 
Then the fraction of volume above a critical density $\rho_c$ is 
\begin{equation}\label{eq: fvgo}
     f_V(\xi > \xi_c)= \int_{\xi_c}^\infty p  d\xi 
                            = \frac{1}{2}  - \frac{1}{2} \text{erf}\Big(\frac{\xi_c-\mu}{\sqrt{2\sigma^2}}\Big),
\end{equation}
where $\xi_c =   \ln (\rho_c / \langle\rho\rangle)$. 
Fig. \ref{fig: lcapp} illustrates $ f_V(\xi > \xi_c)$ as a function of $\xi_c$ with 
$\langle M_s \rangle = 10$ and $b\approx0.5$
\citep{PN02}.
It shows that as a result of shock compressions driven by supersonic turbulent flows, 
strong density enhancements are developed, 
which are spatially clustered and concentrated in a small fraction of the volume
\citep{RoG18}.

These density clumps arising from the compressive motions in supersonic turbulence give rise to a flat isotropic density spectrum
\citep{BLC05}. 
By taking the logarithm of density, which effectively filters out the density peaks and the related compressive component of turbulence, 
\citet{BLC05}
found that 
the density statistics exhibit the same Kolmogorov spectrum and 
scale-dependent anisotropy as the velocity statistics of solenoidal turbulence.
Therefore, by separately studying the turbulence traced by different density structures, we are able to extract the 
different statistical properties corresponding to the solenoidal and compressive components of supersonic turbulence. 
As regards velocity statistics in supersonic turbulence,
with random sampling, the overall velocity spectrum follows the Kolmogorov scaling 
on all length scales as dictated by the dominant solenoidal component of turbulence
\citep{KowL10}.
By selectively sampling the velocity field traced by density peaks, 
which originate from compressive turbulent motions,
the observed spectrum deviates from the Kolmogorov scaling and tends to 
follow the Burgers scaling toward small length scales.
Therefore, 
conditional sampling of turbulent velocities based on the inhomogeneous density distribution 
can be used to separate out the compressional component from the 
solenoidal component in supersonic turbulence.

\section{Application of projected velocity SF to a turbulent MC}

\begin{figure}
\centering
\subfigure[]{
   \includegraphics[width=8.5cm]{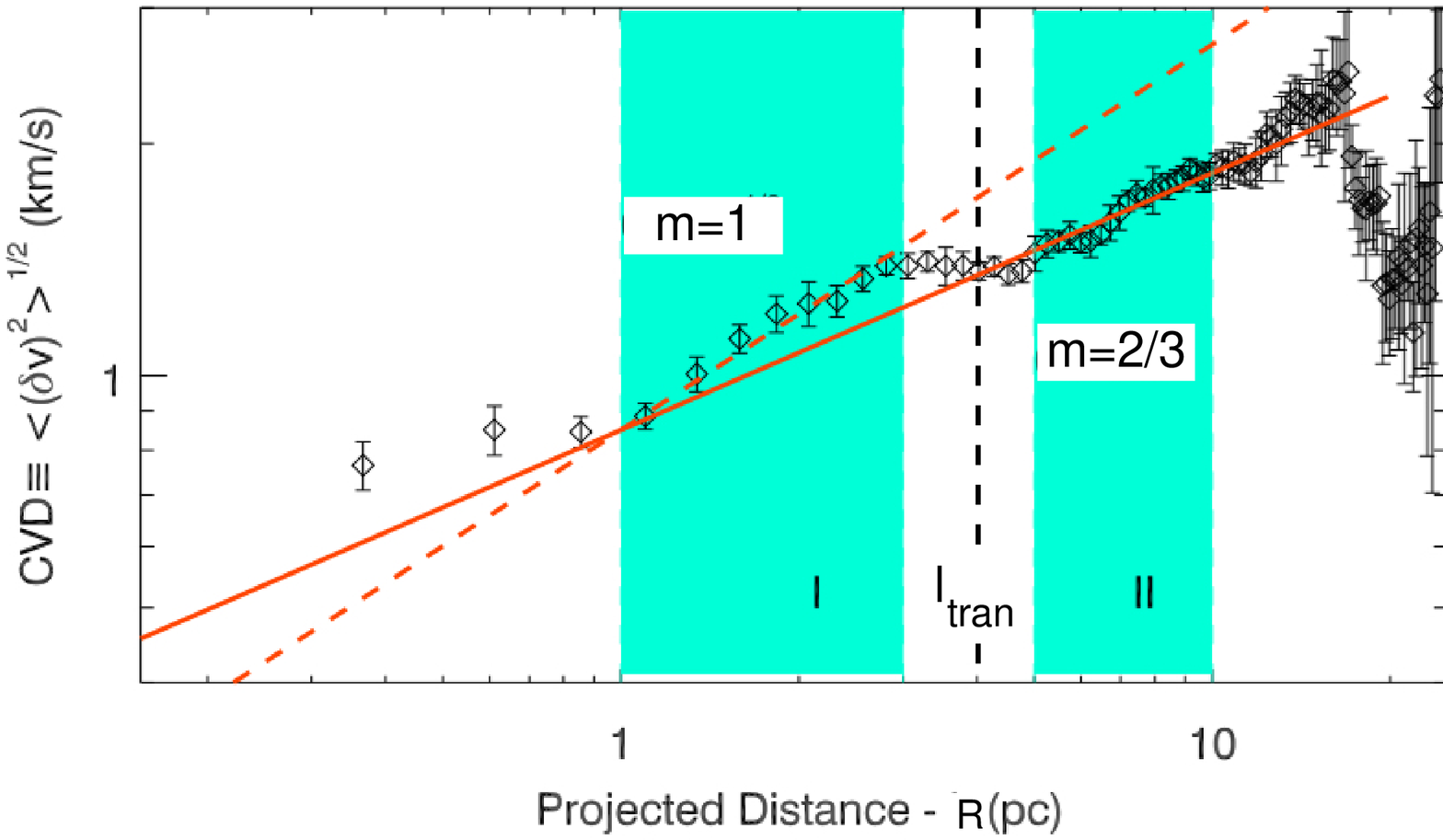}\label{fig: qian}}
\subfigure[]{
   \includegraphics[width=8cm]{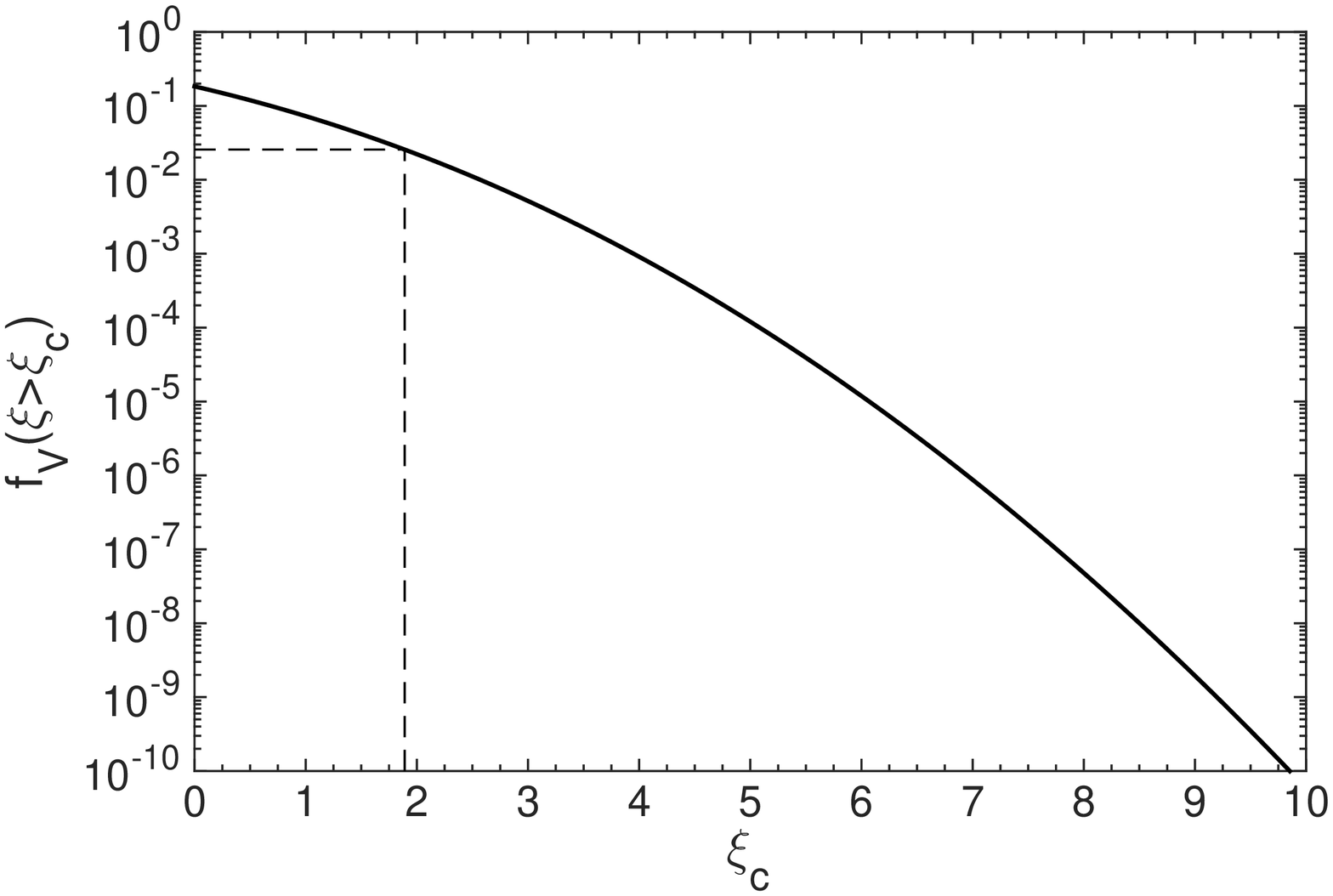}\label{fig: lcapp}}
\caption{ (a)   CVD vs. R for Taurus cloud taken from 
\citet{Qi18}. The added vertical dashed line indicates the transition from the Kolmogorov ($m=2/3$) to Burgers ($m=1$) scaling. 
(b) $f_V(\xi > \xi_c)$ vs. $\xi_c$ in supersonic isothermal turbulence (solid line, Eq. \eqref{eq: fvgo}).
The dashed lines indicate 
the value of $\xi_c$ corresponding to $l_\text{tran}$ shown in (a).  }
\label{fig: obs}
\end{figure}

Turbulence in MCs is highly supersonic 
\citep{Zu74}. 
The core velocity dispersion (CVD) technique
\citep{Qi12}
has been used to probe turbulent velocities in MCs
\citep{Qi15}. 
Molecular cores are treated as point-like tracers
and used to sample the turbulent velocity field at density peaks. 
Only the peak LOS velocities of the emission profiles of cores are adopted, 
and the CVD is defined as the velocity difference between each pair of cores at a certain projected distance between the cores. 
Fig. \ref{fig: qian} is taken from 
\citet{Qi18}
and presents the CVD as a function of projected distance for Taurus cloud. 
As discussed above, 
due to the selective sampling of turbulent velocities by using dense cores, 
the CVD is expected to reveal both the Kolmogorov scaling of solenoidal turbulence and 
the Burgers scaling of compressive turbulence.

By applying the projected velocity SF in \S \ref{sec:sf} to this observational result, 
we compare the scaling of $D(R)^{1/2}$ (Eq. \eqref{eq: drpjt}) with that of the CVD$(R)$, 
where $u_z$ is taken as the peak LOS velocity of a core 
and $R$ is the projected separation distance between a pair of cores. 
The fits to the CVD measurements suggest 
\citep{Qi18}
\begin{subnumcases}
   { \text{CVD}(R) \propto }
    R^\frac{1}{3},     ~~~~~    5 \text{pc} < R < 10 \text{pc}, \\
    R^\frac{1}{2},     ~~~~~    1 \text{pc} < R < 3 \text{pc}, 
\end{subnumcases}
which corresponds to 
\begin{subnumcases}
   { D(R) \propto \label{eq: obsdr}}
    R^\frac{2}{3},     ~~~~~    5 \text{pc} < R < 10 \text{pc}, \\
    R,     ~~~~~~~    1 \text{pc} < R < 3 \text{pc}. 
\end{subnumcases}
The comparison of Eqs. \eqref{eq: drtic} and \eqref{eq: drthc} with the above scaling 
suggests that the cloud is thin with $L < L_i$, 
which is consistent with the finding in 
\citet{Qi15}. 
Further comparison between Eq. \eqref{eq: drtic} and \eqref{eq: obsdr} shows
that the measurements fall in the regime of $L< R < L_i$.

The scalings in Eq. \eqref{eq: obsdr} in the range $L< R < L_i$ indicate 
\begin{subnumcases}
   { m = }
   \frac{2}{3},     ~~~~~    5 \text{pc} < R < 10 \text{pc}, \\
    1,     ~~~~~~    1 \text{pc} < R < 3 \text{pc},
\end{subnumcases}
as shown in Fig. \ref{fig: qian}.          
According to Eq. \eqref{eq: ressf}, there is 
\begin{subnumcases}
   {\alpha = }
   -\frac{11}{3},     ~~~~~    5 \text{pc} < R < 10 \text{pc}, \\
    -4,     ~~~~~~~~    1 \text{pc} < R < 3 \text{pc},
\end{subnumcases}
which corresponds to the Kolmogorov and Burgers scalings, respectively (\S \ref{ssec: sctv}). 
The Kolmogorov spectrum observed on large length scales shows 
the dominance of solenoidal turbulent motions in most of the cloud volume. 
On smaller length scales, the cores preferentially trace the localized dense regions, 
which only occupy a small fraction of the volume and are associated with 
compressive turbulent motions (see Section \ref{ssec: sdf}). 
Therefore, the observed spectrum transits toward the Burgers scaling.
The transition from the Kolmogorov to Burgers scaling occurs at around $l_\text{tran} = 4$ pc (see Fig. \ref{fig: qian}), 
which characterizes the size of a cluster of cores. 
This is consistent with the typical size of a core cluster inferred from the bimodal distribution of core velocity difference found in
\citet{Qi12}. 
$f_V$ corresponding to $l_\text{tran}$ in a thin cloud can be approximately estimated as 
\begin{equation}
    f_V \sim \Big(\frac{l_\text{tran}}{25 ~\text{pc}}\Big)^2 = 0.026.
\end{equation}
Note that 
$25$ pc is the transverse scale of Taurus
\citep{Qi15}.
Accordingly, we have 
$\xi_c \sim 1.89$ and 
$\rho_c / \langle\rho\rangle \sim 6.62$ (see Fig. \ref{fig: lcapp}). 
It means that the cores in Taurus are spatially clustered in localized regions with a typical size of $\sim 4$ pc and 
densities $\rho \gtrsim 6.62 \langle\rho\rangle$,
where the compressive turbulent motions are dominant. 
 
On large scales, the value of $L_i$ can be inferred from the scale where $D(R)$ begins to saturate (see Fig. \ref{fig: thin}). 
We note that CVD measurements are limited to the range of length scales below the transverse size of a cloud, which is $\sim 25$ pc for Taurus
\citep{Qi15}
and can be treated as the lower limit of $L_i$. 
On small scales, the value of $L$ can be inferred from the scale where $D(R)$ begins to steepen due to the projection effect (see Fig. \ref{fig: thin}). 
This is not seen from the CVD measurements of Taurus,
implying that the LOS thickness of core clusters in Taurus
is smaller than the minimum projected separation resolved 
in the observation.

As expected, 
with their physical origin associated with compressive turbulent motions, 
dense cores are useful for targeted sampling of the compressional component of turbulence in an MC, which is 
overall dominated by solenoidal turbulence.

\section{Conclusions}

The SF of projected observables can be generally applied to statistical studies of interstellar turbulence.
Besides rotation measure fluctuations related to turbulent density fluctuations
(LP16, \citealt{XuZ16}), 
here we apply it to projected velocity fluctuations, i.e., $D(R)$ in Eq. \eqref{eq: drpjt}, 
as a direct probe of the dynamics of interstellar turbulence. 
{It enables us to retrieve the underlying spectral form of turbulent velocities 
from the observed 2D SF in different regimes. 
In the case of a thin cloud with the cloud thickness $L$ less than the driving scale of turbulence $L_i$, 
the values of $L$ and $L_i$ can be inferred from the shape of the 2D SF over a broad range of length scales. }

The solenoidal and compressional components of supersonic turbulence have distinctively different velocity scalings. 
Despite its small volume filling factor, 
the latter component can be selectively sampled by dense tracers that are associated with compressive turbulent motions. 
Therefore, 
the velocity SF measured with selective sampling can be used for separately studying different turbulence components over different 
ranges of length scales.

As an example of its application to an MC with selective sampling, we employ the projected velocity SF 
to analyze the CVD measurements of the Taurus cloud by 
\citet{Qi18}. 
As a new technique developed by 
\citet{Qi12}, 
the CVD is less affected by density fluctuations 
compared with other techniques using molecule tracers, 
and provides a relatively clean measurement of turbulent velocities. 
With decreasing length scales, we found a transition from the Kolmogorov scaling of solenoidal turbulent motions to 
the Burgers scaling of compressive turbulent motions. 
It shows that for the supersonic turbulence in Taurus, 
the solenoidal component is dominant across most of the cloud volume.
By contrast, 
the compressional component, which is preferentially sampled by clusters of dense cores, 
is dominant in small-scale isolated regions with the typical size characterized by the transition length scale. 
The small-scale highly compressive regions are found to be coincident with star-forming regions 
in the Orion B MC
\citep{Ork17}. 
It implies that the combination of the projected velocity SF and the CVD provides a useful tool for 
directly probing the gas dynamics, 
separately identifying the solenoidal and compressive components of turbulence on different length scales, and 
characterizing the star-forming regions in MCs.

The projected velocity SF can also be generally used to analyze the 
velocity statistics of other diffuse interstellar phases 
and examine the relation of their scalings and dynamics to those of MCs. 
Its applications to other regimes, including a thin cloud with random sampling and a thick cloud, 
will be explored with future observations. 
\\

S.X. thanks Jianfu Zhang, 
Georgia Panopoulou, 
and Aris Tritsis for stimulating discussions. 
S.X. acknowledges the support for Program number HST-HF2-51400.001-A provided by NASA through a grant from the Space Telescope Science Institute, which is operated by the Association of Universities for Research in Astronomy, Incorporated, under NASA contract NAS5-26555.

\appendix 

\section{Analytical approximations of $D(R)$ in different regimes} 
\label{sec:app}

In the case of a thin cloud with $L < L_i$, 
Eq. \eqref{eq: drpjt} has approximate expressions in different regimes. 
We find 
\begin{equation}\label{eq: app1thin}
\begin{aligned}
   D(R)   &  \approx 4  \langle u_z^2 \rangle \int_0^L d\Delta z (L-\Delta z)  
                   \frac{  (R^2 + \Delta z^2)^\frac{m}{2}  }{L_i^m  } \\
             &  \approx  4  \langle u_z^2 \rangle \int_0^R d\Delta z L  
                   \frac{  R^m  }{L_i^m  }    \\
             &  =   4  \langle u_z^2 \rangle L_i^{-m}  L R^{m+1}     
\end{aligned}
\end{equation}
in the range $R < L < L_i$, 
\begin{equation}\label{eq: app2thin}
\begin{aligned}
   D(R)    & \approx 4  \langle u_z^2 \rangle \int_0^L d\Delta z (L-\Delta z)  
                \frac{ R^m}{L_i^m  }   \\
               & = 2  \langle u_z^2 \rangle  L_i^{-m} L^2  R^m
\end{aligned}
\end{equation}
in the range $L < R < L_i$, and 
\begin{equation}\label{eq: app3thin}
   D(R)     \approx 4  \langle u_z^2 \rangle \int_0^L d\Delta z (L-\Delta z)  
                = 2 \langle u_z^2 \rangle L^2             
\end{equation}
in the range $R > L_i$. 
In the case of a thick cloud with $L > L_i$,
Eq. \eqref{eq: drpjt} is approximately 
\begin{equation}\label{eq: app1thick}
   D(R)     \approx 4  \langle u_z^2 \rangle \int_0^R d\Delta z L  
                   \frac{R^m}{L_i^m  }  
                = 4  \langle u_z^2 \rangle L_i^{-m} L  R^{m+1} ,
\end{equation}
which applies to the inertial range of turbulence, i.e., $R < L_i $.

\bibliographystyle{apj.bst}
\bibliography{xu}

\end{document}